\documentclass[10pt,aps,prd]{revtex4}
\def\be{\begin{equation}}
\def\ee{\end{equation}}
\def\beq{\begin{equation}}
\def\eeq{\end{equation}}
\def\bea{\begin{eqnarray}}
\def\eea{\end{eqnarray}}
\def\no{\nonumber}
\usepackage{amsmath}
\usepackage{relsize}
\usepackage{mathptmx}

\begin{document}

\title{Clebsch-Gordan coefficients for the Higgs algebra: The L\"{o}wdin-Shapiro approach}
\author{T. Shreecharan}
\email{shreecharan@gmail.com}
\affiliation{School of Physics, University of Hyderabad, Hyderabad, Andhra Pradesh, India-500 046}
\begin{abstract}
\begin{center}
{\small ABSTRACT}
\end{center}
The L\"{o}wdin-Shapiro projection operator for the Higgs algebra is constructed and utilised to find an analytical expression for the Clebsch-Gordan coefficients for the same.
\end{abstract}

\maketitle

\section{Introduction}

The coupling of two or more angular momenta to yield a resultant one is accomplished by the vector coupling scheme. In case of two, Clebsch-Gordan coefficients (CGCs) make their appearance and for three momenta the Racah coefficients emerge. This technique can, in principal, be carried out for any number of momentum additions. But the complexity of this naive approach forces one to seek alternatives. It was precisely to tackle this problem that the projection operator (PO) approach was developed by L\"{o}wdin \cite{Low} and later improved upon by Shapiro \cite{Shapiro}. In the aforementioned references it was effectively used to calculate the CGCs for the $su(2)$ algebra. Later, this method has not only been used to find the CGCs of the superanalog of the classical $su(2)$ algebra i.e., $osp(1|2)$ \cite{Berezin}, but also their $q$ deformations namely, $U_q(su(2))$ \cite{Tolstoy} and $U_q(osp(1|2))$ \cite{Minnaert} respectively. This was further used to include the two parameter deformation, $su_{p,q}(2)$, of  the $su(2)$ algebra \cite{Smirnov}. Ref. \cite{Tolstoyrev} contains a useful discussion of POs constructed for various algebras and also their usefulness in the study of representation theory of the same.

In the present work we will construct the PO for the cubic also known as the Higgs algebra (HA):
\be \label{Higgs-alg}
[J_+, J_-] = 2 \beta_1 \, J_0 +  4\beta_2 \, J_0^3, \qquad [J_\pm,~J_\pm]=\pm J_\pm.
\ee
This algebra originally arose as a symmetry algebra in the study of the Kepler problem in curved space, particularly on a sphere \cite{Higgs,Leemon}. It also arises in the quantum mechanical problem of a particle in a Coulomb potential constrained to move on the surface of a sphere. The interesting feature of the above algebra is that it can be viewed as a nonlinear deformation of $su(2)$ or $su(1,1)$ depending on the sign of the linear term. That is, if $\beta_1 = 1$ then the algebra is treated as a deformation of $su(2)$ (Higgs) algebra and if it is equal to $-1$ then it is deformed $su(1,1)$ algebra. It has been shown in \cite{Zhedanov} that the HA is a truncation up to the cubic term of the $su_q(2)$ algebra. In this sense the HA sits between the Lie and $q$ deformed algebras. From the foregoing discussion it can be expected that the representation theory of these algebras is quite interesting and has been studied by a number of authors \cite{Daskaloyannis,Bonatsos,Beckers,Rocek,Quesne,Sunilthese,Bambah}. In the present work we are concerned with the Higgs algebra treated as the deformation of $su(2)$ and therefore are concerned only with the finite dimensional representations (DRs).

The present manuscript is structured in the following manner. In the next section we present the representation theory of cubic angular momentum algebra.  Section III is devoted to the construction of the PO via the L\"{o}wdin-Shapiro technique. Section IV contains its explicit matrix representation. The PO and its matrix representation, derived in the earlier two sections, are utilised in section V to derive an analytical expression for the CGCs. Our conclusions are presented in section VI.

\section{Polynomial $su(2)$ algebra and its representation}

Polynomial deformations of the angular momentum algebra are characterised by the following commutation relations
\be \label{polsu2}
[J_+, J_-] = P(J_0) \equiv g(J_0) - g(J_0-1) \ , \qquad [J_0, J_\pm] = \pm \, J_\pm,
\ee
$g$'s are called the structure functions and their usefulness lies in the fact that they can be used to pin down the Casimir in an almost trivial manner
\be \label{polsu2cas}
\mathcal{J} = \frac{1}{2}\left[\{J_+, J_-\} + g(J_0) + g(J_0-1)\right].
\ee
Similar to the $su(2)$ algebra, the finite DRs are characterised by an integer or half-integer $j$ of dimension $2j~+~1$. By considering a basis in which both the Casimir $\mathcal{J}$ and $J_0$ are diagonal
\be
\mathcal{J} \ \vert j, m  \rangle = g(j) \ \vert j, m  \rangle \qquad J_0 \ \vert j, m  \rangle = m \ \vert j, m  \rangle,
\ee
the action of the ladder operators is as follows
\be
J_+ \, \vert j, m  \rangle  =   \sqrt{g(j) - g(m)} \, \vert j, m+1  \rangle, \qquad
J_- \, \vert j, m  \rangle  =   \sqrt{g(j) - g(m-1)} \, \vert j, m-1  \rangle.
\ee
With an eye on later calculations, an arbitrary state $\vert j, m  \rangle$ is expressed  in terms of a Fock type basis. For this we choose $m=-j+n$, with this identification $\vert j, m  \rangle \equiv \vert j, -j + n  \rangle$. In the new basis the step operators act according to
\bea \label{su2ladfock}
J_0 \, \vert j, n  \rangle  & = &  (-j + n) \, \vert j, n  \rangle \\ \no
J_+ \, \vert j, n  \rangle  & = &   \sqrt{g(j) - g(-j+n)} \, \vert j, n+1  \rangle, \\ \no
J_- \, \vert j, n  \rangle  & = &   \sqrt{g(j) - g(-j+n-1)} \, \vert j, n-1  \rangle.
\eea
In the above equations and in what follows we supress $-j$ dependence in the second entry of the ket vectors. Let us mention here that new finite DRs other than those presented in equation (\ref{su2ladfock}) have been constructed in \cite{Beckers}.

Defining
\be  \label{genfacsu2}
\psi_n = g(j) - g(-j+n-1),
\ee
the ladder operators in (\ref{su2ladfock}) can be written in a more compact form
\be
J_+ \, \vert j, n  \rangle = \sqrt{\psi_{n+1}} \, \vert j, n+1  \rangle \qquad J_- \, \vert j, n  \rangle = \sqrt{\psi_{n}} \, \vert j, n-1  \rangle.
\ee
Thus, a general state $\vert j, n  \rangle$ can be constructed from the ground state in the following manner
\be \label{arbit-state}
J_+^n \vert j, 0 \rangle = \sqrt{[\psi_n]!} \ \vert j, n  \rangle.
\ee
Where we have used the factorial notation of $\psi_n$ to denote
\be
[\psi_n]! = \prod_{k=1}^{n} \psi_k \quad \mathrm{and} \quad [\psi_0]!=1.
\ee
It must be pointed out that the discussion so far is valid for any polynomial deformation of the angular momentum algebra.

In what follows, we will be concerned only with the deformations that are odd degree $2p-1 \ (p=1,2, \cdots)$ for the polynomial;
\be \label{polsu2exp}
P(J_0) = 2 \sum_{r=1}^p \beta_r J_0^r \sum_{s=1}^r (J_0+1)^{r-s} \ (J_0 - 1)^{s-1}.
\ee
$\beta_r$'s are some real non-zero parameters. For a such a polynomial, the structure function is not difficult to find and is
\be
g(J_0) = \sum_{r=1}^p \beta_r [J_0 (J_0+1)]^r.
\ee
Using the above in equation (\ref{genfacsu2}) we get
\be
\psi_n = \sum_{r=1}^p \beta_r \ [j^r(j+1)^r - (j-n+1)^r (j-n)^r ],
\ee
which can be rewritten as
\be \label{genfac2}
\psi_n = n \ (2j+1-n) \ \xi_n, \qquad [\psi_n]! = n! \ [2j+1-n]! \ [\xi_n]! \ ,
\ee
Casting the structure function in the above form has the advantage that the contribution due to the nonlinear terms of the algebra is completely encoded in $\xi_n$, which in the linear limit goes to one. The generalised factorial of $[2j+1-n]!$ is
\be
[2j+1-n]! = (2j)\ (2j-1) \ \cdots \ (2j+1-n) = \frac{(2j)!}{(2j-n)!} = n! \ \left(\begin{array}{c} 2j \\ n \end{array}\right) \ .
\ee
The deformation factor
\be \label{dffacsu2}
\xi_n = \sum_{r=1}^p \sum_{s=1}^r \ \beta_r \ [j(j+1)]^{r-s} \ [(j - n)\ (j - n + 1)]^{s-1},
\ee
which is a polynomial of degree $2p-2$, can be written in a more illuminating manner by factorising it in $n$,
\be
\xi_n = \beta_p \ (n-a_1) \ (n-a_2) \cdots (n-a_{2p-2}),
\ee
so that the generalised factorial of $\xi_n$ becomes
\be \label{dffha}
\big[\xi(a_1, \cdots, a_{2p-2})_n\big]! = \beta_p^n \ (1-a_1)_n \ \cdots \ (1-a_{2p-2})_n \equiv \beta_p^n \prod_{i=1}^{2p-2} (1-a_i)_n.
\ee
In the above $a_i$'s are the roots of $\xi_n=0$ and $(a)_n$ the Pochhammer symbol:
\be
(a)_n = a \ (a+1) \ (a+2) \cdots (a+n-1) = \frac{\Gamma(a+n)}{\Gamma(a)}, \qquad (a)_0 = 1.
\ee
Let us clearly explain our notation. The generalised factorial of the deformation is always given by
\be
\big[\xi(\diamond, \star, \cdots, \bullet, \triangleright)_\# \big]! := \beta_p^\# \ (1-\diamond)_\# \ (1-\star)_\# \ \cdots \ (1-\bullet)_\# (1-\triangleright)_\# \ .
\ee
For later on use let us rewrite Eq. (\ref{arbit-state}) in terms of Eq. (\ref{genfac2})
\be \label{arbit-state-m}
J_+^{j+m} \ \big\vert j, -j \big\rangle = \sqrt{\frac{(j+m)!\ (2j)!}{(j-m)!}} \ \sqrt{\big[\xi(a_1, \cdots, a_{2p-2})_{j+m}\big]!} \ \big\vert j, m  \big\rangle.
\ee
In the present work we will restrict ourselves further and  set $p=2$ and $\beta_1=1$ in Eq. (\ref{polsu2exp}) so that the polynomial is cubic
\be
P(J_0) = 2 J_0 + 4 \beta_2 J_0^3,
\ee
and the deformation factor, Eq. (\ref{dffacsu2}), turns out to be
\be \label{dffacHsu2}
\xi_n = 1 + \beta_2 \ \Big[n^2 - (2j+1) \ n + 2j\ (j+1) \Big].
\ee
Hermiticity requirement of the step operators yields $\beta_2 \geq - 1/2j^2$. The expression corresponding to Eq. (\ref{arbit-state-m}) takes the form
\be \label{arbit-state-ha}
J_+^{j+m} \ \big\vert j, -j \big\rangle = \sqrt{\frac{(j+m)!\ (2j)!}{(j-m)!}\ \big[\xi(a_+,a_-)_{j+m} \big]!}  \ \big\vert j, m  \big\rangle ,
\ee
with $\big[\xi(a_+,a_-)_{j+m} \big]!$ being a special case of Eq. (\ref{dffha}). The roots of Eq. (\ref{dffacHsu2}) are
\be \label{Hrootssu2}
a_\pm = \frac{1}{2} \left[(2j+1) \pm \sqrt{(2j+1)^2 - 8 j(j+1) - 4/\beta_2 } \right].
\ee
Now that we have discussed the representation theory corresponding to the HA in a notation that will be useful for us, we immediately proceed to the next section wherein the PO is constructed.

\section{The projection operator}

In this section we will derive the PO, denoted by $\mathcal{P}^j$. It acts linearly in the space $\mathcal{V}$, the direct sum of all representation spaces $\mathcal{V}_j$. The following requirement is imposed on $\mathcal{P}^j$:
\be \label{cond1}
\mathcal{P}^j \ \big\vert j, j  \big\rangle  =  \big\vert j, j  \big\rangle.
\ee
To ensure that the projected vector is an eigenstate of $\mathcal{J}$ and $J_0$ we further impose the additional conditions
\bea \label{cond2}
J_0 \ \mathcal{P}^j \ \big\vert j, j  \big\rangle & = & j \ \big\vert j, j  \big\rangle, \\ \label{cond3}
J_+ \ \mathcal{P}^j \ \big\vert j, j  \big\rangle & = & 0.
\eea
Similar to the L\"{o}wdin-Shapiro approach \cite{Low,Shapiro}, the ansatz for the PO is taken to be of the form
\be \label{rawproj}
\mathcal{P}^j = \sum_{\ell=0}^{\infty} C_{\ell} \ J_-^\ell \  J_+^\ell.
\ee
The exponents are same in the above ansatz since $[\mathcal{P}^j,J_0]=0$, which in turn is a consequence of Eq. (\ref{cond2}). Now all that remains to be done is to fix the constants $C_\ell$. Substituting Eq. (\ref{rawproj}) in Eq. (\ref{cond3}) we get
\be \label{rawproj2}
\sum_{\ell=0}^{\infty} C_{\ell} \ J_+ \ J_-^\ell \  J_+^\ell \ \big\vert j, j  \big\rangle = 0.
\ee
Our aim will be to shift the $J_+$ operator to the right of $J^\ell_-$. The relevant formula is
\be \label{shiftform}
J_+ \ J^\ell_- =  J^\ell_- \ J_+ + 2 \ J^{\ell-1}_- \sum_{r=1}^{\ell} \ (J_0+1-r) + 4 \beta_2 \ J^{\ell-1}_- \sum_{r=1}^{\ell} \ (J_0+1-r)^3 .
\ee
Using the above relation in Eq. (\ref{rawproj2}) results in
\bea \no
\sum_{\ell=0}^{\infty} C_{\ell} \ J_-^\ell \ J_+^{\ell+1} \ \big\vert j, j  \big\rangle + \sum_{\ell=1}^{\infty} \ell \ C_{\ell} \ J_-^{\ell-1} \ (2 J_0 - \ell + 1) \ J_+^{\ell} \ \big\vert j, j  \big\rangle \\
+ \beta_2 \ \sum_{\ell=1}^{\infty} \ell \ C_{\ell} \ J_-^{\ell-1} \ (2 J_0 - \ell + 1) \ \big[\ell^2 - (2 J_0+1) \ell + 2 J_0 \ (J_0+1)\big] \ J_+^{\ell} \ \big\vert j, j  \big\rangle = 0.
\eea
We have simplified the summations of Eq. (\ref{shiftform}) to obtain the above expression. Redefining $\ell \rightarrow \ell-1$ in the first summation and making use of the general formula $J_0^q \  J_\pm^{p} = J_\pm^{p} \ (J_0 \pm p)^q$, so that $J_0$ can be replaced by its eigenvalue. The above equation can then be cast as
\bea \no
\sum_{\ell=1}^{\infty} C_{\ell-1} \ J_-^{\ell-1} \ J_+^{\ell} \ \big\vert j, j  \big\rangle + \sum_{\ell=1}^{\infty} \ \ell \ (2 j + \ell + 1) \ C_{\ell}  \ J_-^{\ell-1} \ J_+^{\ell} \ \big\vert j, j  \big\rangle  \\
+ \beta_2 \ \sum_{\ell=1}^{\infty} \ \ell \ (2 j + \ell + 1) \ \big[\ell^2 + (2 j +1) \ \ell + 2 j \ (j+1) \big] \ C_{\ell} \ J_-^{\ell-1} \ J_+^{\ell} \ \big\vert j, j  \big\rangle = 0,
\eea
which, in turn, is obeyed if the coefficients follow the recurrence relation
\be
\ell \ (2 j + \ell + 1) \ \Big\{1 + \beta_2 \ \big[\ell^2 + (2 j +1) \ \ell + 2 j \ (j+1) \big] \Big\} \ C_{\ell} + C_{\ell-1} = 0,
\ee
for $\ell = 1, 2, 3, \cdots$. The solution of the above relation is given by
\bea \no
C_{0} = (-1)^\ell \ell! \ (2j+2)_\ell \ \big[\xi(b_+,b_-)_\ell \big]! \ C_\ell,
\eea
with
\be \label{roots}
b_\pm = \frac{1}{2} \left[-(2j+1) \pm \sqrt{(2j+1)^2 - 8 j(j+1) - 4/\beta_2 } \right].
\ee
Now that $C_\ell$ has been determined, from equation (\ref{rawproj}) we get
\be \label{rawproj3}
\mathcal{P}^{j} = C_{0} \ \mathlarger{\sum}_{\ell=0}^{\infty} \ \frac{(-1)^\ell \ J_-^\ell \ J_+^\ell}{\ell! \ (2j+2)_\ell \ \big[\xi (b_+,b_-)_\ell\big]!}.
\ee
The constant $C_0$ can be fixed by multiplying the above equation on both sides from the left by $\big\vert j, j  \big\rangle$. Making use of Eq. (\ref{cond1}) for the left-hand-side and noticing that only $\ell=0$ term contributes to the right-hand-side, leads to $C_0=1$.

We have derived the projection operator for the special case of the highest weight state only. This is not of much utility unless it can be extended to an arbitrary state. Such an operator is given by
\be \label{rawproj4}
\mathcal{P}_{jm} = \mathcal{N} \ J_-^{j-m} \ \mathcal{P}^{j} \ J_+^{j-m}.
\ee
The arbitrary constant $\mathcal{N}$ can be determined by a condition analogous to Eq. (\ref{cond1}): $P_{jm} \big\vert j, m \big\rangle  = \big\vert j, m \big\rangle$ leading to
\be
\mathcal{N} \ \big\langle j, m \big\vert \ J_-^{j-m} \ J_+^{j-m} \ \big\vert j, m \big\rangle = 1.
\ee
The matrix element is not difficult to calculate and when the job is done it leads to the following value for the constant
\be \label{normalisation}
\mathcal{N} = \frac{(j+m)!}{(j-m)! \ (2j!)} \ \frac{\big[\xi(a_+,a_-)_{j+m} \big]!}{\big[\xi(a_+,a_-)_{2j} \big]!} \ .
\ee
The final form of the PO for an arbitrary state can be obtained by using equations (\ref{rawproj3}) and (\ref{normalisation}) in Eq. (\ref{rawproj4}):
\be \label{finalproj}
\mathcal{P}_{jm} = \frac{(2j+1)\ (j+m)!}{(j-m)!} \ \frac{\big[\xi(a_+,a_-)_{j+m} \big]!}{\big[\xi(a_+,a_-)_{2j} \big]!} \
\mathlarger{\sum}_{\ell=0}^{\infty} \ \frac{(-1)^\ell}{\ell!} \frac{ J_-^{j-m+\ell} \ J_+^{j-m+\ell} }{(2j+\ell+1)! \ \big[\xi(b_+,b_-)_\ell \big]!}.
\ee
The entire derivation can be carried out by starting from the lowest state as well. The expression for this PO can be obtained by replacing $m \rightarrow -m$ and interchanging $J_-$ and $J_+$ in the above;
\be \label{finalproj-low}
\mathcal{P}_{jm} = \frac{(2j+1)\ (j-m)!}{(j+m)!} \ \frac{\big[\xi(a_+,a_-)_{j-m} \big]!}{\big[\xi(a_+,a_-)_{2j} \big]!} \
\mathlarger{\sum}_{\ell=0}^{\infty} \ \frac{(-1)^\ell}{\ell!} \frac{ J_+^{j+m+\ell} \ J_-^{j+m+\ell} }{(2j+\ell+1)! \ \big[\xi(b_+,b_-)_\ell \big]!}.
\ee
We are now in a position to derive the CG coefficients for the HA. For that we will need the matrix element of the PO, which we explicitly calculate in the next section.

\section{Matrix representation of the projection operator}

Let us consider the matrix element
\be \label{matp1}
\mathcal{M} = \Big\langle \prod_{i=1}^N j_i m^\prime_i \Big\vert \ \mathcal{P}_{jm} \ \Big\vert \prod_{i=1}^N j_i m_i \Big\rangle .
\ee
where we have
\be
\Big\vert \prod_{i=1}^N j_i m_i \Big\rangle = \prod_{i=1}^N \big\vert j_i m_i \big\rangle.
\ee
Note that the above is a simultaneous eigenstate of $J_i^2$, $J_{i0}$, and $J_0$ with the eigenvalues given by $j_i$, $m_i$, and $m = \sum_i^N m_i$ respectively. Thus Eq. (\ref{matp1}) has non zero elements only when
\be
\sum_i^N m_i = \sum_i^N m^\prime_i = m \ .
\ee
Using the expression of the PO, Eq. (\ref{finalproj}) of the previous section, in equation (\ref{matp1}) we get
\be \label{int-proj}
\mathcal{M} = \frac{(2j+1)\ (j+m)!}{(j-m)!} \ \frac{\big[\xi(a_+,a_-)_{j+m} \big]!}{\big[\xi(a_+,a_-)_{2j} \big]!} \
\mathlarger{\sum}_{\ell=0}^{\infty} \ \frac{(-1)^\ell}{\ell!} \frac{ \mathcal{M}^\ell }{(2j+\ell+1)! \ \big[\xi(b_+,b_-)_\ell \big]!}.
\ee
Where
\be \label{marr}
\mathcal{M}^\ell = \big\langle j_i m^\prime_i \big\vert \ J_-^{j-m+\ell} \ J_+^{j-m+\ell} \ \big\vert j_i m_i \big\rangle \ .
\ee
Note that the product symbol has been suppressed for the sake of convenience. We will simplify $\mathcal{M}^\ell$ by setting $J_\pm = \sum_{i=1}^N J_{i \pm}$ and using the multinomial theorem:
\be
\big( J_{1+} + \cdots + J_{N+} \big)^t = \mathlarger{\sum}_{t_1 \cdots t_N} \ \left(\begin{array}{c} t \\ t_1 \cdots t_N \end{array}\right) \ J_{1+}^{\ t_1} \cdots J_{N+}^{\ t_N} \ ,
\ee
where
\be
\left(\begin{array}{c} t \\ t_1 \cdots t_N \end{array}\right) = \frac{t !}{t_1! \cdots t_N!} \ ,
\ee
with the constraint that $t_1+ \cdots + t_N = t$, equation (\ref{marr}) can be cast as
\be \label{mat-ele}
\mathcal{M}^\ell = [(j-m+\ell)!]^2 \mathlarger{\sum}_{s_1 \cdots s_N} \ \mathlarger{\sum}_{t_1 \cdots t_N} \ \left(\ \mathlarger{\prod}_{i=1}^N \ \frac{\big\langle j_i m^\prime_i \big\vert \ J_{i-}^{\ s_i} \ J_{i+}^{\ t_i} \ \big\vert j_i m_i \big\rangle}{(s_i)! \ (t_i)!} \right).
\ee
Where the factorial squared appearing outside the summations is via the identification $\sum s_i = \sum t_i = j-m+\ell$. The limits of the sums given above are
\be
0 \leq s_i \leq j_i - m_i^\prime \ , \qquad 0 \leq t_i \leq j_i - m_i \ .
\ee
It can be noticed that the nonzero elements of Eq. (\ref{mat-ele}) occur when $m_i^\prime + s_i = m_i + t_i$. This fact can be utilised in eliminating one set of summations either over $t$'s or $s$. This is achieved by introducing
\be
k_i = j_i - m_i^\prime - s_i = j_i - m_i - t_i \ ,
\ee
and having the limits
\be
0 \leq  k_i \leq \mathrm{min} \ \big(j_i - m_i^\prime,j_i - m_i \big)\ .
\ee
They also satisfy the relation
\be
\sum_{i=1}^N k_i = \sum_{i=1}^N \big( j_i - m_i - t_i \big) = \mathbf{j} - j - \ell\ ,
\ee
$\mathbf{j}$ being the maximum possible $j$ that arises when adding individual angular momenta. With these simplifications we get
\be \label{mat-element1}
\mathcal{M}^\ell = [(j-m+\ell)!]^2 \mathlarger{\sum}_{k_1 \cdots k_N} \left[\mathlarger{\prod}_{i=1}^N \frac{\big\langle j_i m^\prime_i \big\vert \ J_{i-}^{j_i - m_i^\prime - k_i} \ J_{i+}^{j_i - m_i - k_i} \ \big\vert j_i m_i \big\rangle}{(j_i - m_i^\prime - k_i)! \ (j_i - m_i - k_i)!} \right] \ .
\ee
Our task will be to explicitly calculate $\mathcal{M}^\ell$. As a first step let us define
\be \label{fks1}
f_{k_i}^i = \frac{\big\langle j_i m^\prime_i \big\vert \ J_{i-}^{j_i - m_i^\prime - k_i} \ J_{i+}^{j_i - m_i - k_i} \ \big\vert j_i m_i \big\rangle}{(j_i - m_i^\prime - k_i)! \ (j_i - m_i - k_i)!} \ .
\ee
Making use of the action of the ladder operators on the states via Eq. (\ref{arbit-state-ha}) we get
\be
f_{k_i}^i = \left[\left(\begin{array}{c} 2 j_i \\ j_i - m^\prime_i \end{array}\right) \ \left(\begin{array}{c} 2 j_i \\ j_i - m_i \end{array}\right) \right]^{1/2} \ \frac{(j_i - m^\prime_i)! \ (j_i - m_i)! \  (2 j_i - k_i)!}{(j_i - m^\prime_i-k_i)! \ (j_i - m_i-k_i)! \ (2j_i)! \ (k_i)!} \ \frac{\big[\xi(a_+,a_-)_{2j_i-k_i} \big]!}{\sqrt{\big[\xi(a_+,a_-)_{j_i+m^\prime_i} \big]! \big[\xi(a_+,a_-)_{j_i+m_i} \big]!}} \ .
\ee
It can be shown that
\be
\big[\xi(a_+,a_-)_{2j_i-k_i} \big]! = \frac{\big[\xi(a_+,a_-)_{2j_i} \big]!}{\big[\xi(A_+,A_-)_{k_i} \big]!} \ ,
\ee
with $A_\pm = (1-a_\pm + 2 j_i)$. It follows from the above two equations that
\be \label{fkzero}
f_{0}^i = \left[\left(\begin{array}{c} 2 j_i \\ j_i - m^\prime_i \end{array}\right) \ \left(\begin{array}{c} 2 j_i \\ j_i - m_i \end{array}\right) \right]^{1/2} \ \frac{\big[\xi(a_+,a_-)_{2j_i} \big]!}{\sqrt{\big[\xi(a_+,a_-)_{j_i+m^\prime_i} \big]! \big[\xi(a_+,a_-)_{j_i+m_i} \big]!}} \ .
\ee
Therefore $f_{k_i}^i$ can be written as a product of terms that are independent of $k_i$, namely $f_{0}^i$, and those that are not $\mathbf{P}_{k_i}$. Explicitly
\be
\mathbf{P}_{k_i} = \frac{(-1)^{k_i}}{(k_i)!} \ \frac{[-(j_i-m_i)]_{k_i} \ [-(j_i-m^\prime_i)]_{k_i}}{(-2j_i)_{k_i} \big[\xi(A_+,A_-)_{k_i} \big]!}
\ee
with these observations we have
\be \label{melement-2}
\mathcal{M}^\ell = [(j-m+\ell)!]^2 \ \Big(\prod_{i=1}^N \ f_0^i \Big) \ \sum_{k_1 \cdots k_N} \ \prod_{i=1}^N \ \mathbf{P}_{k_i} \ .
\ee
The final expression for the matrix element that we are seeking is obtained by substituting Eq. (\ref{melement-2}) in Eq. (\ref{int-proj}):
\be
\mathcal{M} = \frac{(2j+1)\ (j+m)!}{(j-m)!} \ \frac{\big[\xi(a_+,a_-)_{j+m} \big]!}{\big[\xi(a_+,a_-)_{2j} \big]!} \
\Big( \prod_{i=1}^N \ f_0^i \Big) \ \mathlarger{\sum}_{\ell=0}^{\infty} \ \frac{(-1)^\ell}{\ell!} \frac{[(j-m+\ell)!]^2}{(2j+\ell+1)! \ \big[\xi(b_+,b_-)_\ell \big]!} \ \sum_{k_1 \cdots k_N} \ \prod_{i=1}^N \ \mathbf{P}_{k_i}.
\ee
%

\section{Nonlinear CG coefficients}

In the following section we show how the PO can be used to calculate the CG coefficients. We will work out explicitly the case of addition of two angular momenta: $\mathbf{J}_1 + \mathbf{J}_2 = \mathbf{J}$. The coupled state vector is denoted in the usual form $\vert j_1 j_2 \ j m \rangle$. The PO can be constructed from this basis in the following manner
\be
\mathcal{P}_{jm} = \big\vert j_1 j_2 \ j m \big\rangle \big\langle j m \ j_1 j_2 \big \vert \ .
\ee
Equation (\ref{matp1}) for $i=2$ becomes
\be
\mathcal{M} = \big\langle j_2 m^\prime_2  \ j_1 m^\prime_1 \big \vert \ \mathcal{P}_{jm} \ \big\vert j_1 m_1 \ j_2 m_2 \big\rangle =
\langle j_2 m^\prime_2  \ j_1 m^\prime_1 \big\vert j_1 j_2 \ j m \big\rangle \big\langle j m \ j_1 j_2 \big\vert j_1 m_1 \ j_2 m_2 \big\rangle \ .
\ee
Performing some algebraic manipulations the formula for the CGCs can be written in the form
\be \label{proj-final}
\big\langle j m \ j_1 j_2 \big\vert j_1 m_1 \ j_2 m_2 \big\rangle = \frac{\big\langle j_2 m^\prime_2  \ j_1 m^\prime_1 \big\vert \ \mathcal{P}_{jm} \ \big\vert j_1 m_1 \ j_2 m_2 \big\rangle}{\sqrt{\big\langle j_2 m^\prime_2  \ j_1 m^\prime_1 \big\vert \ \mathcal{P}_{jm} \ \big\vert j_1 m^\prime_1 \ j_2 m^\prime_2 \big\rangle}} \ .
\ee
Setting $m^\prime_1 = j_1$ and using the constraint $m_1 + m_2 = m^\prime_1 + m^\prime_2 = m$ we get
\be \label{proj-final2}
\big\langle j m \ j_1 j_2 \big\vert j_1 m_1 \ j_2 m_2 \big\rangle = \frac{\big\langle j_2 m-j_1  \ j_1 j_1 \big\vert \ \mathcal{P}_{jm} \ \big\vert j_1 m_1 \ j_2 m_2 \big\rangle}{\sqrt{\big\langle j_2 m-j_1  \ j_1 j_1 \big\vert \ \mathcal{P}_{jm} \ \big\vert j_1 j_1 \ j_2 m-j_1 \big\rangle}} \equiv \frac{\mathcal{P}_{jm}^{Num}}{\sqrt{\mathcal{P}_{jm}^{Den}}} \ .
\ee
The result for the numerator is
\bea \no
\mathcal{P}_{jm}^{Num} & = & \left[\frac{(2j+1)^2 \ [(j+m)!]^2 \ (2j_1)! \ (j_2-m_2)! \ (j_1+j_2-m)!}{[(j-m)!]^2 (j_1+m_1)!(j_1-m_1)!(j_2+m_2)!(j_2-j_1+m)!} \frac{[\xi(a_+,a_-)_{j+m}]!^2 \ [\xi(a_+,a_-)_{2j_1}]!}{[\xi(a_+,a_-)_{2j}]!^2 \ [\xi(a_+,a_-)_{j_1+m_1}]!} \right. \\ \no
&& \left. \frac{[\xi(a_+,a_-)_{2j_2}]!^2}{[\xi(a_+,a_-)_{j_2+m_2}]! \ [\xi(a_+,a_-)_{j_2-j_1+m}]!} \right]^{1/2} \\ \label{fnum}
&& \mathlarger{\sum}_{\substack{\ell, \ k \\ \ell+k=j_1+j_2-j}} \ \frac{(-1)^\ell}{(\ell)!} \ \frac{[(j-m+\ell)!]^2 }{(2j+\ell+1)!} \frac{1}{[\xi(b_+,b_-)_{\ell}]!} \frac{1}{(k)!} \frac{(2j_2-k)!}{(j_1+j_2-m-k)!\ (j_2-m_2-k)!} \frac{1}{[\xi(A_+,A_-)_{k}]!} \ .
\eea
which can be simplified further to the final form
\bea \no
\mathcal{P}_{jm}^{Num} & = & \left[\frac{(2j+1)^2 \ [(j+m)!]^2 \ (2j_1)! \ (j_2-m_2)! \ (j_1+j_2-m)!}{[(j-m)!]^2 (j_1+m_1)!(j_1-m_1)!(j_2+m_2)!(j_2-j_1+m)!} \frac{[\xi(a_+,a_-)_{j+m}]!^2 \ [\xi(a_+,a_-)_{2j_1}]!}{[\xi(a_+,a_-)_{2j}]!^2 \ [\xi(a_+,a_-)_{j_1+m_1}]!} \right. \\ \no
&& \left. \frac{[\xi(a_+,a_-)_{2j_2}]!^2 \ [\xi(b_+,b_-)_{j_2+j_1-j}]!^{-2}}{[\xi(a_+,a_-)_{j_2+m_2}]! \ [\xi(a_+,a_-)_{j_2-j_1+m}]!} \right]^{1/2} \ (-1)^{j_1+j_2-j}\\ \label{fnum}
&& \mathlarger{\sum}_{k} \ \frac{(-1)^k}{(k)!} \ \frac{(j_1+j_2-m-k)! \ (2j_2-k)! }{(j_1+j_2-j-k)! (j_1+j_2+j+1-k)!(j_2-m_2-k)! } \frac{[\xi(B_+,B_-)_{k}]!}{[\xi(A_+,A_-)_{k}]!} \ .
\eea
Where $B_\pm = (1-b_\pm + j_1 + j_2 - j)$. The above sum runs over the values
\be
0 \leq k \leq \textrm{min}(j_2-m_2,j_1+j_2-j) \ .
\ee
The denominator can be obtained by setting $m_1=j_1$ and $m_2=m-j_1$, after $k=j_1+j_2-j-\ell$ in the above
\bea \no
\mathcal{P}_{jm}^{Den} & = & \left[\frac{(2j+1) (j+m)! (j_1+j_2-m)!}{(j-m)! \ (j_2-j_1+m)!} \frac{[\xi(a_+,a_-)_{j+m}]! \ [\xi(a_+,a_-)_{2j_2}]!}{[\xi(a_+,a_-)_{2j}]! \ [\xi(a_+,a_-)_{j_2-j_1+m}]!} \frac{1}{[\xi(b_+,b_-)_{j_2+j_1-j}]!} \right] \\
&& \mathlarger{\sum}_{\ell} \ \frac{(-1)^{\ell}}{\ell !} \ \frac{(j_2-j_1+j+\ell)!}{(2j+\ell+1)!(j_2+j_1-j-\ell)!} \frac{[\xi(B_+,B_-)_{j_1+j_2-j-\ell}]!}{[\xi(A_+,A_-)_{j_1+j_2-j-\ell}]!}
 \ .
\eea
The summation in the above equation can be cast as a hypergeometric function thereby leading to the final result
\bea \no
\mathcal{P}_{jm}^{Den} & = & \left[\frac{(2j+1) (j+m)! (j_1+j_2-m)!}{(j-m)! \ (j_2-j_1+m)! (2j+1)!} \frac{(j_2-j_1+j)!}{(j_2+j_1-j)!} \frac{[\xi(a_+,a_-)_{j+m}]! \ [\xi(a_+,a_-)_{2j_2}]!}{[\xi(a_+,a_-)_{2j}]! \ [\xi(a_+,a_-)_{j_2-j_1+m}]!} \frac{1}{[\xi(b_+,b_-)_{j_2+j_1-j}]!} \right. \\ \label{fden} && \left. \frac{[\xi(B_+,B_-)_{j_2+j_1-j}]!}{[\xi(A_+,A_-)_{j_2+j_1-j}]!} \right] \ {_4}F_3\big(j-j_1-j_2,1+j-j_1+j_2,1-\tilde{A}_+,1-\tilde{A}_-;2j+2,1-b_+,1-b_-;1\big).
\eea
Wherein $\tilde{A}_\pm = (1-a_\pm + j_2-j_1+j)$ and ${_4}F_3$ is the generalised hypergeometric series. The final result corresponding to Eq. (\ref{proj-final2}) is now easy to obtain. Dividing Eq. (\ref{fnum}) with the square-root of Eq. (\ref{fden}) leads to final answer for the nonlinear CG coefficients
\bea \no
&& \big\langle j m \ j_1 j_2 \big\vert j_1 m_1 \ j_2 m_2 \big\rangle = (-1)^{j_1+j_2-j}  \Bigg[\frac{(2j+1)(2j+1)!(2j_1)!(j_2-m_2)!(j+m)!}{(j_1+m_1)!(j_1-m_1)!(j_2+m_2)!(j-m)!}  \frac{(j_2+j_1-j)!}{(j_2-j_1+j)!} \frac{1}{{_4}F_3} \\
&& \frac{[\xi(a_+,a_-)_{j+m}]!}{[\xi(a_+,a_-)_{2j}]!} \frac{[\xi(a_+,a_-)_{2j_1}]!\  [\xi(a_+,a_-)_{2j_2}]!}{[\xi(a_+,a_-)_{j_1+m_1}]!\ [\xi(a_+,a_-)_{j_2+m_2}]!} \frac{[\xi(A_+,A_-)_{j_1+j_2-j}]!}{[\xi(B_+,B_-)_{j_1+j_2-j}]!}  \Bigg]^{1/2} \\ \no
&& \mathlarger{\sum}_{k} \ \frac{(-1)^{k}}{(k)!} \ \frac{(j_1+j_2-m-k)! \ (2j_2-k)! }{(j_1+j_2-j-k)! (j_1+j_2+j+1-k)!(j_2-m_2-k)! } \frac{[\xi(B_+,B_-)_{k}]!}{[\xi(A_+,A_-)_{k}]!}
\eea
\section{Conclusions}

We have shown that it is possible to obtain the L\"{o}wdin-Shapiro projection operator for the Higgs algebra and have used it to find the Clebsch-Gordan type coefficients for the same. It is known that the usual $su(2)$ coefficients are related to the discrete polynomials it would be interesting to know what polynomials these coefficients give rise to.

\acknowledgements

The author thanks UGC-India for financial support through the Dr. D. S. Kothari Post-Doctoral Fellowship Scheme.


\begin{thebibliography}{99}

\bibitem{Low} P.-O. L\"{o}wdin, \textit{Angular momentum wavefunctions constructed by projector operators}, Rev. Mod. Phys. \textbf{36} (1964) 966.

\bibitem{Shapiro} J. Shapiro, \textit{Matrix reperesentation of the angular momentum projection operator}, J. Math. Phys. \textbf{6} (1965) 1680.

\bibitem{Berezin} F. A. Berezin and V. N. Tolstoy, \textit{The group with Grassmann structure $UOSP(1|2)$}, Commun. Math. Phys. \textbf{78} (1981) 409.

\bibitem{Tolstoy} Yu. F. Smirnov, V. N. Tolstoy, and Yu. I. Kharitonov, \textit{Method of projection operators and $q$-analog of the quantum angular momentum theory. I. Clebsch-Gordan coefficients and irreducible tensor operators}, Sov. J. Nucl. Phys. \textbf{53} (1991) 593; \textit{Projection-operator method and the $q$-analog of the quantum theory of angular momentum Racah coefficients, $3j-$ and $6j-$ symbols and their symmetry propetrties}, Sov. J. Nucl. Phys. \textbf{53} (1991) 1068.

\bibitem{Minnaert} P. Minnaert and M. Mozrzymas, \textit{Clebsch-Gordan coefficients for the quantum superalgebra $U_q(osp(1|2))$}, J. Math. Phys. \textbf{35} (1994) 3132; P. Minnaert and M. Mozrzymas, \textit{Properties of Clebsch-Gordan coefficients and $3-j$ symbols for the quantum superalgebra $U_q(osp(1|2))$}, J. Phys. \textbf{A 28} (1995) 669.

\bibitem{Smirnov} Yu. F. Smirnov and R. F. Wehrhahn, \textit{Clebsch-Gordan coefficients for the two paramter quantum superalgebra $SU_{p,q}(2)$ in the L\"{o}wdin-Shapiro approach}, J. Phys. \textbf{A 25} (1992) 5563.

\bibitem{Tolstoyrev} V. N. Tolstoy, \textit{Fortieth anniversary of extremal projector method for Lie symetries}, Contemporary Mathematics \textbf{391} (2005) 391.

\bibitem{Higgs} P. W. Higgs, \textit{Dynamical symmetries in a spherical geometry. I}, J. Phys. \textbf{A 12} (1979) 309.

\bibitem{Leemon} H. I. Leemon, \textit{Dynamical symmetries in a spherical geometry. II}, J. Phys. \textbf{A 12} (1979) 489.

\bibitem{Zhedanov} A. S. Zhedanov, \textit{The Higgs algebra as a quantum deformation of $SU(2)$}, Mod. Phys. Lett. \textbf{A 7} (1992) 507.


\bibitem{Daskaloyannis} C. Daskaloyannis, \textit{Generalized deformed oscillator and nonlinear algebras}, J. Phys. \textbf{A 26} (1993) 26 L871.

\bibitem{Bonatsos} D. Bonatsos, C. Daskaloyannis, and P. Kolokotronis, \textit{Generalized deformed $SU(2)$ algebra}, J. Phys. \textbf{A 24} (1991) L789.

\bibitem{Beckers} B. Abdesselam, J. Beckers, A. Chakrabarti, and N. Debergh, \textit{On nonlinear angular momentum theories, their representations and associated Hopf structures}, J. Phys. \textbf{A 29} (1996) 3075. 

\bibitem{Quesne} C. Delbecq and C. Quesne, \textit{Nonlinear deformations of $su(2)$ and $su(1,1)$ generalizing Witten's algebra}, J. Phys. \textbf{A 26} (1993) L127.

\bibitem{Rocek} M. Ro\"{c}ek, \textit{Representation theory of the nonlinear $SU(2)$ algebra}, Phys. Lett. \textbf{B 255} (1991) 554.

\bibitem{Sunilthese} V. Sunilkumar, \textit{Aspects of polynomial algebras and their physical applications}, thesis in University of Hyderabad, math-ph/0203047.

\bibitem{Bambah} V. Sunilkumar, B. A. Bambah, and R. Jagannathan, \textit{Jordan–Schwinger-type realizations of three-dimensional polynomial algebras}, Mod. Phys. Lett. \textbf{A 17} (2002) 1559.


\end{thebibliography}
\end{document}